%% file: Cattaneo-Titiunik-VazquezBare_2020_Stata.tex
\documentclass[bib]{statapress}

\usepackage[crop,newcenter,frame]{pagedims}

\usepackage{sj}
\usepackage{bm}

\usepackage{stata}

\usepackage{shadow}

\usepackage{hyperref}
\hypersetup{colorlinks=true,       
    linkcolor=blue,          
    citecolor=blue,        
    filecolor=magenta,      
    urlcolor=blue           
}

\usepackage{mathrsfs}

\sjsetissue{$vv$}{$ii$}{$mm$}{$xxxx$}

\usepackage{verbatim}
\usepackage{graphicx}
\usepackage{amssymb,amsmath,bbm,longtable}
\usepackage{subcaption}
\usepackage{float}

\renewcommand{\P}{\mathbbm{P}}
\newcommand{\E}{\mathbb{E}}

\newcommand{\1}{\mathbbm{1}}

\begin{document}

\inserttype[st0001]{article}
\author{M. D. Cattaneo, R. Titiunik and G. Vazquez-Bare}{
  Matias D. Cattaneo\\Princeton University\\Princeton, NJ\\cattaneo@princeton.edu
  \and
  Rocio Titiunik\\Princeton University\\Princeton, NJ\\titiunik@princeton.edu
  \and
  Gonzalo Vazquez-Bare\\UC Santa Barbara\\Santa Barbara, CA\\gvazquez@econ.ucsb.edu
}

\title[RD Designs with Multiple Cutoffs or Scores]{Analysis of Regression Discontinuity Designs with Multiple Cutoffs or Multiple Scores}

\maketitle

\begin{abstract}
We introduce the \texttt{Stata} (and \texttt{R}) package \texttt{rdmulti}, which includes three commands (\texttt{rdmc}, \texttt{rdmcplot}, \texttt{rdms}) for analyzing Regression Discontinuity (RD) designs with multiple cutoffs or multiple scores. The command \texttt{rdmc} applies to non-cumulative and cumulative multi-cutoff RD settings. It calculates pooled and cutoff-specific RD treatment effects, and provides robust bias-corrected inference procedures. Post estimation and inference is allowed. The command \texttt{rdmcplot} offers RD plots for multi-cutoff settings. Finally, the command \texttt{rdms} concerns multi-score settings, covering in particular cumulative cutoffs and two running variables contexts. It also calculates pooled and cutoff-specific RD treatment effects, provides robust bias-corrected inference procedures, and allows for post-estimation estimation and inference. These commands employ the \texttt{Stata} (and \texttt{R}) package \texttt{rdrobust} for plotting, estimation, and inference. Companion \texttt{R} functions with the same syntax and capabilities are provided.\medskip


\end{abstract}

\begin{center} \bigskip This version: \today \end{center}

\newpage

\section{Introduction}

Regression discontinuity (RD) designs with multiple cutoffs or multiple scores are commonly encountered in empirical work in  Economics, Education, Political Science, Public Policy, and many other disciplines. As a consequence, these specific settings have also received attention in the recent RD methodological literature \citep*[][and references therein]{Papay-Willett-Murnane_2011_JoE,Reardon-Robinson_2012_JREE,Wong-Steiner-Cook_2012-JEBS,Keele-Titiunik_2015_PA,Keele-Titiunik-Zubizarreta_2015_JRSSA,Cattaneo-Keele-Titiunik-VazquezBare_2016_JOP,Cattaneo-Keele-Titiunik-VazquezBare_2021_JASA}. In this article, we introduce the software package \texttt{rdmulti}, which includes three \texttt{Stata} commands (and analogous \texttt{R} functions) for the analysis of RD designs with multiple cutoffs or multiple scores.

The command \texttt{rdmc} applies to non-cumulative and cumulative multi-cutoff RD settings, following recent work in \citet*{Cattaneo-Keele-Titiunik-VazquezBare_2016_JOP,Cattaneo-Keele-Titiunik-VazquezBare_2021_JASA}. Specifically, it calculates pooled and cutoff-specific RD treatment effects, employing local polynomial estimation and robust bias-corrected inference procedures. Post estimation and inference is allowed. The companion command \texttt{rdmcplot} offers RD plots for multi-cutoff settings. Finally, the command \texttt{rdms} concerns multi-score settings, covering in particular cumulative cutoffs and bivariate score contexts. It also calculates pooled and cutoff-specific RD treatment effects based on local polynomial methods, and allows for post-estimation estimation and inference. These commands employ the \texttt{Stata} (and \texttt{R}) package \texttt{rdrobust} for plotting, estimation, and inference; see \citet*{Calonico-Cattaneo-Titiunik_2014_Stata,Calonico-Cattaneo-Titiunik_2015_R,Calonico-Cattaneo-Farrell-Titiunik_2017_Stata} for software details. See also \citet*{Cattaneo-Titiunik-VazquezBare_2017_JPAM} for a comparison of RD methodologies, \citet*{Cattaneo-Idrobo-Titiunik_2019_Book,Cattaneo-Idrobo-Titiunik_2020_Book} and \citet*{Cattaneo-Titiunik-VazquezBare_2020_Sage} for a practical introductions to RD designs, and \citet{Cattaneo-Escanciano_2017_AIE} for a recent edited volume with further references.

To streamline the presentation, this article employs only simulated data to showcase all three settings covered by the package \texttt{rdmulti}: non-cumulative multiple cutoffs, cumulative multiple cutoffs, and bivariate score settings. For further discussion and illustration employing real data sets see \citet*{Cattaneo-Idrobo-Titiunik_2020_Book}. The three settings covered by the package correspond, respectively, to (i) RD designs where different subgroups in the data are exposed to distinct but only one of the cutoff points (non-cumulative case), (ii) RD designs where units receive one single score and units are confronted to a sequence of ordered cutoffs points (cumulative case), and (iii) RD designs where units received two scores and there is a boundary on the plane determining the control and treatment areas. Well-known examples of each of these settings are:\smallskip\newline
$\bullet$ \textit{Non-Cumulative Multiple Cutoffs}: units in different groups (e.g., schools) receive an univariate score (e.g., test score) but the RD cutoff varies by group;\smallskip\newline
$\bullet$ \textit{Cumulative Multiple Cutoffs}: units receive the an univariate score (e.g., age) but different treatments are assigned at distinct score levels (e.g., at age 60 and at age 65);\smallskip\newline
$\bullet$ \textit{Multiple Scores}: units receive two scores (e.g., latitude and longitude) and treatment is assigned based on a boundary depending on both scores (e.g., geographic boundary).

We elaborate further on these cases in the upcoming sections, where we also give graphical representations of each case.

The \texttt{Stata} (and \texttt{R}) package \texttt{rdmulti} complements several recent software packages for RD designs. First, it explicitly relies on \texttt{rdrobust} \citep*{Calonico-Cattaneo-Titiunik_2014_Stata,Calonico-Cattaneo-Titiunik_2015_R,Calonico-Cattaneo-Farrell-Titiunik_2017_Stata} for implementation, and hence further extends its scope to the case of RD designs with multiple cutoffs or multiple scores. Second, while the package focuses on local polynomial methods, related methods employing local randomization ideas and implemented in the package \texttt{rdlocrand} can also be used in the contexts of multiple cutoffs and multiple scores \citep*{Cattaneo-Titiunik-VazquezBare_2016_Stata}. Third, the package \texttt{rddensity} \citep*{Cattaneo-Jansson-Ma_2018_Stata} can also be used in multiple cutoffs or multiple scores settings for falsification purposes. Finally, see the package \texttt{rdpower} \citep*{Cattaneo-Titiunik-VazquezBare_2018_rdpower} for power calculations and sampling design methods, which can also be applied in the contexts discussed in this article.

The rest of the article is organized as follows. Section \ref{section:methods} gives a brief overview of the methods implemented in the package \texttt{rdmulti}, and also provides further references. Sections \ref{section:rdmc}, \ref{section:rdmcplot} and \ref{section:rdms} discuss the syntax of the commands \texttt{rdmc}, \texttt{rdmcplot} and \texttt{rdms}, respectively. Section \ref{section:illustration} gives numerical illustrations, and Section \ref{section:conclusion} concludes. The latest version of this software, as well other software and materials useful for the analysis of RD designs, can be found at:\newline
\begin{center}\vspace{-.3in}\href{https://sites.google.com/site/rdpackages/}{\texttt{https://sites.google.com/site/rdpackages/}}\end{center}

\section{Overview of Methods}\label{section:methods}

In this section we briefly describe the main ideas and methods used in the package \texttt{rdmulti}. For further methodological details see \citet{Keele-Titiunik_2015_PA}, \cite*{Cattaneo-Keele-Titiunik-VazquezBare_2016_JOP,Cattaneo-Keele-Titiunik-VazquezBare_2021_JASA}, \cite*{Cattaneo-Idrobo-Titiunik_2020_Book}, and references therein. All estimation and inference procedures employ rdplots \citep*{Calonico-Cattaneo-Titiunik_2015_JASA} as well as local polynomial point estimation and robust bias correction inference methods \citep*{Calonico-Cattaneo-Titiunik_2014_ECMA,Calonico-Cattaneo-Farrell_2018_JASA,Calonico-Cattaneo-Farrell-Titiunik_2019_RESTAT,Calonico-Cattaneo-Farrell_2020_ECTJ,Calonico-Cattaneo-Farrell_2020_wp}.

\subsection{Non-cumulative Multiple Cutoffs}

In this case, individuals have a running variable $X_i$ and a vector of potential outcomes $(Y_i(0),Y_i(1))$. Each individual faces a cutoff $C_i\in\mathcal{C}$ with $\mathcal{C}=\{c_1,c_2,\ldots c_J\}$. For example, \citet{Chay-etal_2005} study the effect of a school improvement program introduced in 1990 by the Chilean government. In this program, low-performing schools received public funding to improve infrastructure and teacher training, among other things. Assignment to this program was based on a school-level measure of test scores falling below a cutoff, where the cutoff was different across Chile's 13 administrative regions. In this example, $C_i$ indicates each school's administrative region, since this determines the cutoff faced by each school.

Unlike in a standard single-cutoff RD design, $C_i$ is a random variable. In a sharp design, individuals are treated when their running variable exceeds their corresponding cutoff, $D_i=\1(X_i\ge C_i)$. A key feature of this design is that the variable $C_i$ partitions the population, that is, each unit faces one and only one value of $C_i$. As the notation suggests, the potential outcomes for each individual are the same regardless of the specific cutoff they are exposed to; see \cite*{Cattaneo-Keele-Titiunik-VazquezBare_2016_JOP,Cattaneo-Keele-Titiunik-VazquezBare_2021_JASA} for more discussion. Finally, we only consider finite multiple cutoffs because this is the most natural setting for empirical work: in practice, continuous cutoff are discretized for estimation and inference, as discussed and illustrated below.

Under regularity conditions, which include smoothness of conditional expectations among other things (see aforementioned references for details), the cutoff-specific treatment effects, $\tau(c)=\E[Y_i(1)-Y_i(0)|X_i=c,C_i=c]$, are identified by:

\begin{equation}\label{eq:cs_effects}
\tau(c)=\lim_{x \downarrow c} \E[Y_i | X_i=x,C_i=c]-\lim_{x \uparrow c} \E[Y_i | X_i=x,C_i=c]
\end{equation}

The pooled RD estimate is obtained by recentering the running variable, $\tilde{X}_i=X_i-C_i$, thus normalizing the cutoff at zero:

\begin{equation}\label{eq:pooled}
\tau_\mathtt{P}= \lim_{x \downarrow 0} \E[Y_i | \tilde{X}_i=x]-\lim_{x \uparrow 0} \E[Y_i | \tilde{X}_i=x],
\end{equation}
where
\begin{equation}
\tau_\mathtt{P} = \sum_{c\in\mathcal{C}} \tau(c) \; \omega(c), \qquad \omega(c)=\frac{f_{X|C}(c|c)\P[C_i=c]}{\sum\limits_{c\in\mathcal{C}} f_{X|C}(c|c)\P[C_i=c]}
\end{equation}

All these parameters can be readily estimated using local polynomial methods \citep*[see][for a practical introduction]{Cattaneo-Idrobo-Titiunik_2019_Book}, conditioning on cutoffs when appropriate. In other words, RD methods can by applied to each cutoff separately, in addition to pooling the data. Therefore, the \texttt{rdmulti} package implements bandwidth selection, estimation and inference based on local polynomial methods using the \texttt{rdrobust} command, described in \citet{Calonico-Cattaneo-Titiunik_2014_Stata,Calonico-Cattaneo-Titiunik_2015_R,Calonico-Cattaneo-Farrell-Titiunik_2017_Stata}. Specifically, the command \texttt{rdmc} allows for multi-cutoffs RD designs.

For the pooled parameter $\tau_\mathtt{P}$, the weights are estimated using the fact that $\omega(c)=\P[C_i=c|\tilde{X}_i=0]$; see \cite*{Cattaneo-Keele-Titiunik-VazquezBare_2016_JOP} for further details. Then, given a bandwidth $h>0$,
\[\hat{\omega}(c)=\frac{\sum_i \1(C_i=c,-h \le \tilde{X}_i\le h)}{\sum_i \1(-h\le \tilde{X}_i \le h)}.\]
When not specified by the user, the \texttt{rdmc} command uses the bandwidth selected by \texttt{rdrobust} when estimating the pooled effect to estimate the weights. 

\subsection{Cumulative Multiple Cutoffs}

In an RD setting with cumulative cutoffs, individuals receive different treatments (or different dosages of a treatment) for different ranges of the running variable. In such setting, individuals receive treatment 1 if $X_i< c_1$, treatment 2 if $c_1\le X_i< c_2$, and so on, until the last treatment value at $X_i\ge c_J$. For example, \citet{Brollo-etal_2013_AER} examine the effect of federal transfers on political corruption in Brazilian municipalities. The amount of the federal transfer that municipalities receive depends on the municipality's population, and changes discretely at specified cutoffs. For example, municipalities with population below 10,189 receive a certain amount, municipalities with population between 10,189 and 13,585 receive a larger amount, and so on.

Denote the values of these treatments as $d_j$, so that the treatment variable is now $D_i\in\{d_1,d_2,\ldots d_J\}$. Under standard regularity conditions, we have:
\begin{align*}
\tau_j=\E[Y_i(d_j)-Y_i(d_{j-1})|X_i=c_j]=\lim_{x \downarrow c_j} \E[Y_i | X_i=x]-\lim_{x \uparrow c_j} \E[Y_i | X_i=x]
\end{align*}

Since, unlike the case with multiple non-cumulative cutoffs, the population is not partitioned, each observation can be used to estimate two different (but contiguous on the score dimension) treatment effects. For example, units receiving treatment dosage $d_j$ are used as ``treated'' (i.e. above the cutoff $c_j$) when estimating $\tau_j$ and as ``controls'' when estimating $\tau_{j+1}$ (i.e. below the cutoff $c_{j+1}$). As a result, cutoff-specific estimators may not be independent, although the dependence disappears asymptotically as long as the bandwidths around each cutoff decrease with the sample size. On the other hand, bandwidths can be chosen to be non-overlapping to ensure that observations are used only once.

Once the data has been assigned to each cutoff under analysis, local polynomial methods can also be applied cutoff by cutoff in the cumulative multiple cutoffs case. We illustrate this approach below; for further discussion see \citet*{Cattaneo-Idrobo-Titiunik_2020_Book}, and the references therein.

\subsection{Multiple Scores}

In a multi-score RD design, treatment is assigned based on multiple running variables and some function determining a treatment ``region'' or ``area''. We focus on the case with two running variables, $\mathbf{X}_i=(X_{1i},X_{2i})$, which is by far the most common case in empirical work. This case occurs naturally when, for instance, a treatment is assigned based on scores in two different exams (such as language and mathematics). \citet{Matsudaira_2008_JoE} estimates the effect of a mandatory summer school program assigned to students who fail to score higher than a preset cutoff in both math and reading exams. Another common case of multiple running variables occurs when a treatment is assigned based or based on geographic location (e.g., latitude and longitude). \citet{Keele-Titiunik_2015_PA} discuss the effect of political campaign advertising on voter turnout and political attitudes by comparing voters in adjacent media markets, which result in different levels of exposure to advertising. 

This type of assignment defines a continuum of treatment effects over the boundary of the treatment region, denoted by $\mathcal{B}$. For instance, if treatment is assigned to students scoring below 50 in language and mathematics, the treatment boundary is $\mathcal{B}=\{x_1\le 50,x_2=50\}\cup \{x_1=50,x_2\le 50\}$. For each point $\mathbf{b}\in\mathcal{B}$, the treatment effect at that point is given by 
\[\tau(\mathbf{b})=\E[Y_i(1)-Y_i(0)|\mathbf{X}_i=\mathbf{b}],\]
and under regularity conditions, 
\[\tau(\mathbf{b})=\lim_{\substack{d(\mathbf{x},\mathbf{b})\to 0, \\ \mathbf{x}\in\mathcal{B}_t}}\E[Y_i|\mathbf{X}_i=\mathbf{x}]-\lim_{\substack{d(\mathbf{x},\mathbf{b})\to 0, \\ \mathbf{x}\in\mathcal{B}_c}}\E[Y_i|\mathbf{X}_i=\mathbf{x}]\]
where $\mathcal{B}_c$ and $\mathcal{B}_t$ denote the control and treatment areas, respectively, and $d(\cdot,\cdot)$ is a metric.

Since estimating a whole curve of treatment effects may not be feasible in practice, it is common to define a set of boundary points of interest at which to estimate the RD treatment effects. In the previous example, for instance, three points of interest on the boundary determining treatment assignment could be $\{(25,50),(50,50),(50,25)\}$. On the other hand, the pooled RD estimand requires defining some measure of distance to the cutoff, such as the perpendicular (Euclidean) distance. This distance can be seen as the recentered running variable $\tilde{X}_i$, which allows defining the pooled estimand as in Equation \ref{eq:pooled}.

\section{\texttt{rdmc} syntax}\label{section:rdmc}

This section describes the syntax of the command \texttt{rdmc}, which estimates the pooled and cutoff-specific RD effects using \texttt{rdrobust}.

\subsection{Syntax}

\begin{stsyntax}
\qquad{
\texttt{rdmc} {\it depvar} {\it runvar} \optif  \optin
    , \underline{c}var({\it cutoff\_var}) \optional{fuzzy({\it string})  \\
    \qquad\qquad \underline{deriv}var({\it string}) pooled\_opt({\it string}) verbose \underline{p}var({\it string}) \\
    \qquad\qquad \underline{q}var({\it string}) \underline{h}var({\it string}) \underline{hr}ightvar({\it string}) \underline{b}var({\it string}) \\
    \qquad\qquad \underline{br}ightvar({\it string}) \underline{rho}var({\it string}) \underline{covs}var({\it string}) \underline{covsdrop}var({\it string}) \\
    \qquad\qquad \underline{kernel}var({\it string}) \underline{weights}var({\it string}) \underline{bwselect}var({\it string}) \\
    \qquad\qquad \underline{scalepar}var({\it string}) \underline{scaleregul}var({\it string}) \underline{masspoints}var({\it string}) \\
    \qquad\qquad \underline{bwcheck}var({\it string}) \underline{bwrestrict}var({\it string}) \underline{stdvars}var({\it string}) \\
    \qquad\qquad \underline{vce}var({\it string}) level({\it \#}) plot graph\_opt({\it string}) }
}
\end{stsyntax}
\color{black}
\hangpara {\it depvar} is the dependent variable.

\hangpara {\it runvar} is the running variable (a.k.a. score or forcing variable).

\hangpara \texttt{\underline{c}var({\it cutoff\_var})} specifies the numeric variable \textit{cutoff\_var} that indicates the cutoff faced by each unit in the sample.

\hangpara \texttt{fuzzy({\it string})} indicates a fuzzy design. See \texttt{help rdrobust} for details.

\hangpara \texttt{\underline{deriv}var({\it string})}  a variable of length equal to the number of different cutoffs that specifies the order of the derivative for \texttt{rdrobust} to calculate cutoff-specific estimates. See \texttt{help rdrobust} for details.

\hangpara \texttt{pooled\_opt({\it string})} specifies the options to be passed to \texttt{rdrobust} to calculate pooled estimate. See \texttt{help rdrobust} for details.

\hangpara \texttt{verbose} displays the output from \texttt{rdrobust} to calculate pooled estimand.

\hangpara \texttt{\underline{p}var({\it string})} a variable of length equal to the number of different cutoffs that specifies the order of the polynomials for \texttt{rdrobust} to calculate cutoff-specific estimates. See \texttt{help rdrobust} for details.

\hangpara \texttt{\underline{q}var({\it string})} a variable of length equal to the number of different cutoffs that specifies the order of the polynomials for bias estimation for \texttt{rdrobust} to calculate cutoff-specific estimates. See \texttt{help rdrobust} for details.

\hangpara \texttt{\underline{h}var({\it string})} a variable of length equal to the number of different cutoffs that specifies the bandwidths for \texttt{rdrobust} to calculate cutoff-specific estimates. When \texttt{hrightvar} is specified, \texttt{hvar} indicates the bandwidth to the left of the cutoff.  When \texttt{hrightvar} is not specified, the same bandwidths are used at each side.  See \texttt{help rdrobust} for details.

\hangpara \texttt{\underline{hright}var({\it string})} a variable of length equal to the number of different cutoffs that specifies the bandwidths to the right of the cutoff for \texttt{rdrobust} to calculate cutoff-specific estimates. When \texttt{hrightvar} is not specified, the same bandwidths in \texttt{hvar} are used at each side.  See \texttt{help rdrobust} for details.

\hangpara \texttt{\underline{b}var({\it string})} a variable of length equal to the number of different cutoffs that specifies the bandwidths for the bias for \texttt{rdrobust} to calculate cutoff-specific estimates. When \texttt{brightvar} is specified, \texttt{bvar} indicates the bandwidth to the left of the cutoff.  When \texttt{brightvar} is not specified, the same bandwidths are used at each side.  See \texttt{help rdrobust} for details.

\hangpara \texttt{\underline{bright}var({\it string})} a variable of length equal to the number of different cutoffs that specifies the bandwidths to the right of the cutoff for \texttt{rdrobust} to calculate cutoff-specific estimates. When \texttt{brightvar} is not specified, the same bandwidths in \texttt{bvar} are used at each side.  See \texttt{help rdrobust} for details.

\hangpara \texttt{\underline{rho}var({\it string})}  a variable of length equal to the number of different cutoffs that specifies the value of rho for \texttt{rdrobust} to calculate cutoff-specific estimates. See \texttt{help rdrobust} for details.

\hangpara \texttt{\underline{covs}var({\it string})}  a variable of length equal to the number of different cutoffs that specifies the covariates for \texttt{rdrobust} to calculate cutoff-specific estimates. See \texttt{help rdrobust} for details.

\hangpara \texttt{\underline{covsdrop}var({\it string})} a variable of length equal to the number of different cutoffs that specifies whether collinear covariates should be dropped. See \texttt{help rdrobust} for details.

\hangpara \texttt{\underline{kernel}var({\it string})} a variable of length equal to the number of different cutoffs that specifies the kernels for \texttt{rdrobust} to calculate cutoff-specific estimates. See \texttt{help rdrobust} for details.

\hangpara \texttt{\underline{weights}var({\it string})} a variable of length equal to the number of different cutoffs that specifies the weights for \texttt{rdrobust} to calculate cutoff-specific estimates. See \texttt{help rdrobust} for details.

\hangpara \texttt{\underline{bwselect}var({\it string})} a variable of length equal to the number of different cutoffs that specifies the bandwidth selection method for \texttt{rdrobust} to calculate cutoff-specific estimates. See \texttt{help rdrobust} for details.

\hangpara \texttt{\underline{scalepar}var({\it string})} a variable of length equal to the number of different cutoffs that specifies the value of scalepar for \texttt{rdrobust} to calculate cutoff-specific estimates. See \texttt{help rdrobust} for details.

\hangpara \texttt{\underline{scaleregul}var({\it string})} a variable of length equal to the number of different cutoffs that specifies the value of scaleregul for \texttt{rdrobust} to calculate cutoff-specific estimates. See \texttt{help rdrobust} for details.

\hangpara \texttt{\underline{masspoints}var({\it string})}  a variable of length equal to the number of different cutoffs that specifies how to handle repeated values in the running variable. See \texttt{help rdrobust} for details.

\hangpara \texttt{\underline{bwcheck}var({\it string})} a variable of length equal to the number of different cutoffs that specifies the value of \texttt{bwcheck}. See \texttt{help rdrobust} for details.

\hangpara \texttt{\underline{bwrestrict}var({\it string})} variable of length equal to the number of different cutoffs that specifies whether computed bandwidths are restricted to the range of \textit{runvar}. See \texttt{help rdrobust} for details.

\hangpara \texttt{\underline{stdvars}var({\it string})} a variable of length equal to the number of different cutoffs that specifies whether \textit{depvar} and \textit{runvar} are standardized. See \texttt{help rdrobust} for details.

\hangpara \texttt{\underline{vce}var({\it string})} a variable of length equal to the number of different cutoffs that specifies the variance-covariance matrix estimation method for \texttt{rdrobust} to calculate cutoff-specific estimates. See \texttt{help rdrobust} for details.

\hangpara \texttt{level({\it \#})} specifies the confidence levels for confidence intervals. See \texttt{help rdrobust} for details.

\hangpara \texttt{plot} plots the pooled and cutoff-specific estimates and the weights given by the pooled estimate to each cutoff-specific estimate.

\hangpara \texttt{graph\_opt({\it string})} options to be passed to the graph when \texttt{plot} is specified.

\section{\texttt{rdmcplot} syntax}\label{section:rdmcplot}

This section describes the syntax of the command \texttt{rdmcplot}, which plots the regression functions for each of the groups facing each cutoff using \texttt{rdplot}.

\subsection{Syntax}

\begin{stsyntax}
\qquad{
\texttt{rdmcplot} {\it depvar} {\it runvar} \optif  \optin
    , \underline{c}var({\it cutoff\_var})  \optional{\underline{nbins}var({\it string}) \\
    \qquad\qquad \underline{nbinsright}var({\it string}) \underline{binselect}var({\it string}) \\
    \qquad\qquad \underline{scale}var({\it string}) \underline{scaleright}var({\it string}) \underline{support}var({\it string}) \\
    \qquad\qquad \underline{supportright}var({\it string}) \underline{p}var({\it string}) \underline{h}var({\it string}) \underline{hright}var({\it string}) \\
    \qquad\qquad \underline{kernel}var({\it string}) \underline{weights}var({\it string}) \underline{covs}var({\it string}) \\
    \qquad\qquad  \underline{covseval}var({\it string}) \underline{covsdrop}var({\it string}) \underline{binsopt}var({\it string})  \\
    \qquad\qquad \underline{lineopt}var({\it string}) \underline{xlineopt}var({\it string}) ci({\it cilevel}) nobins nopoly \\
    \qquad\qquad noxline nodraw genvars}
}
\end{stsyntax}
\color{black}
\hangpara {\it depvar} is the dependent variable.

\hangpara {\it runvar} is the running variable (a.k.a. score or forcing variable).

\hangpara \texttt{\underline{c}var({\it cutoff\_var})} specifies the numeric variable \textit{cutoff\_var} that indicates the cutoff faced by each unit in the sample.

\hangpara \texttt{\underline{nbins}var({\it string})} a variable of length equal to the number of different cutoffs that specifies the number of bins for \texttt{rdplot}. When \texttt{nbinsrightvar} is specified, \texttt{nbinsvar} indicates the number of bins to the left of the cutoff.  When \texttt{nbinsrightvar} is not specified, the same number of bins is used at each side. See \texttt{help rdplot} for details.

\hangpara \texttt{\underline{nbinsright}var({\it string})} a variable of length equal to the number of different cutoffs that specifies the number of bins to the right of the cutoff for \texttt{rdplot}. When \texttt{nbinsrightvar} is not specified, the number of bins in \texttt{nbinsvar} used at each side. See \texttt{help rdplot} for details.

\hangpara \texttt{\underline{binselect}var({\it string})} a variable of length equal to the number of different cutoffs that specifies the bin selection method for \texttt{rdplot}. See \texttt{help rdplot} for details.

\hangpara \texttt{\underline{scale}var({\it string})} a variable of length equal to the number of different cutoffs that specifies the scale for \texttt{rdplot}. When \texttt{scalerightvar} is specified, \texttt{scalevar} indicates the scale to the left of the cutoff. When \texttt{scalerightvar} is not specified, the same scale is used at each side. See \texttt{help rdplot} for details.

\hangpara \texttt{\underline{scaleright}var({\it string})} a variable of length equal to the number of different cutoffs that specifies the scale to the right of the cutoff for \texttt{rdplot}. When \texttt{scalerightvar} is not specified, the scale in \texttt{scalevar} is used at each side. See \texttt{help rdplot} for details.

\hangpara \texttt{\underline{support}var({\it string})} a variable of length equal to the number of different cutoffs that specifies the support for \texttt{rdplot}. When \texttt{supportrightvar} is specified, \texttt{supportvar} indicates the support to the left of the cutoff. When \texttt{supportrightvar} is not specified, the same support is used at each side. See \texttt{help rdplot} for details.

\hangpara \texttt{\underline{supportright}var({\it string})} a variable of length equal to the number of different cutoffs that specifies the support to the right of the cutoff for \texttt{rdplot}. When \texttt{supportrightvar} is not specified, the support in \texttt{supportvar} is used at each side. See \texttt{help rdplot} for details.

\hangpara \texttt{\underline{p}var({\it string})} a variable of length equal to the number of different cutoffs that specifies the order of the polynomials for \texttt{rdplot}. See \texttt{help rdplot} for details.

\hangpara \texttt{\underline{h}var({\it string})} a variable of length equal to the number of different cutoffs that specifies the bandwidths for \texttt{rdplot}. When \texttt{hrightvar} is specified, \texttt{hvar} indicates the bandwidth to the left of the cutoff. When \texttt{hrightvar} is not specified, the same bandwidth is used at each side. See \texttt{help rdplot} for details.

\hangpara \texttt{\underline{hright}var({\it string})} a variable of length equal to the number of different cutoffs that specifies the bandwidth to the right of the cutoff for \texttt{rdplot}. When \texttt{hrightvar} is not specified, the bandwidth in \texttt{hvar} is used at each side. See \texttt{help rdplot} for details.

\hangpara \texttt{\underline{kernel}var({\it string})} a variable of length equal to the number of different cutoffs that specifies the kernels for \texttt{rdplot}. See \texttt{help rdplot} for details.

\hangpara \texttt{\underline{weights}var({\it string})} a variable of length equal to the number of different cutoffs that specifies the weights for \texttt{rdplot}. See \texttt{help rdplot} for details.

\hangpara \texttt{\underline{covs}var({\it string})} a variable of length equal to the number of different cutoffs that specifies the covariates for \texttt{rdplot}. See \texttt{help rdplot} for details.

\hangpara \texttt{\underline{covseval}var({\it string})} a variable of length equal to the number of different cutoffs that specifies the evaluation points for additional covariates. See \texttt{help rdplot} for details.

\hangpara \texttt{\underline{covsdrop}var({\it string})} a variable of length equal to the number of different cutoffs that specifies whether collinear covariates should be dropped. See \texttt{help rdplot} for details.

\hangpara \texttt{\underline{binsopt}var({\it string})} a variable of length equal to the number of different cutoffs that specifies options for the bins plots.

\hangpara \texttt{\underline{lineopt}var({\it string})} a variable of length equal to the number of different cutoffs that specifies options for the polynomial plots.

\hangpara \texttt{\underline{xlineopt}var({\it string})} a variable of length equal to the number of different cutoffs that specifies options for the vertical lines indicating the cutoffs.

\hangpara \texttt{ci({\it cilevel})} adds confidence intervals of level{\it cilevel} to the plot.

\hangpara \texttt{nobins} omits the bins plot.

\hangpara \texttt{nopoly} omits the polynomial curve plot.

\hangpara \texttt{noxline} omits the vertical lines indicating the cutoffs.

\hangpara \texttt{nodraw} omits the plot.

\hangpara \texttt{genvars} generates variables to replicate plots by hand. Variable labels indicate the corresponding cutoff.

\texttt{rdmcplot\_hat\_y\_}{\it c} predicted value of the outcome variable given by the global polynomial estimator in cutoff number {\it c}.

\texttt{rdmcplot\_mean\_x\_}{\it c} sample mean of the running variable within the corresponding bin for each observation in cutoff number {\it c}.

\texttt{rdmcplot\_mean\_y\_}{\it c} sample mean of the outcome variable within the corresponding bin for each observation in cutoff number {\it c}.

\texttt{rdmcplot\_ci\_l\_}{\it c} lower end value of the confidence interval for the sample mean of the outcome variable within the corresponding bin for each observation in cutoff number {\it c}.

\texttt{rdmcplot\_ci\_r\_}{\it c} upper end value of the confidence interval for the sample mean of the outcome variable within the corresponding bin for each observation in cutoff number {\it c}.

\section{\texttt{rdms} syntax}\label{section:rdms}

This section describes the syntax of the command \texttt{rdms}, which analyzes RD designs with cumulative cutoffs or two running variables.

\subsection{Syntax}

\begin{stsyntax}
\qquad{
\texttt{rdms} {\it depvar} {\it runvar1} [{\it runvar2 treatvar}] \optif  \optin
    , \underline{c}var({\it cutoff\_var1} \\
    \qquad\qquad {[cutoff\_var2]}) 
    \optional{range({\it range1 [range2]}) xnorm({\it string}) \\
    \qquad\qquad  fuzzy({\it string}) \underline{deriv}var({\it string}) pooled\_opt({\it string}) \underline{p}var({\it string}) \\
    \qquad\qquad \underline{q}var({\it string}) \underline{h}var({\it string}) \underline{hr}ightvar({\it string}) \underline{b}var({\it string}) \\
    \qquad\qquad \underline{br}ightvar({\it string}) \underline{rho}var({\it string}) \underline{covs}var({\it string}) \underline{covsdrop}var({\it string})  \\
    \qquad\qquad \underline{kernel}var({\it string}) \underline{weights}var({\it string}) \underline{bwselect}var({\it string}) \\
    \qquad\qquad \underline{scalepar}var({\it string}) \underline{scaleregul}var({\it string})  \underline{masspoints}var({\it string}) \\
    \qquad\qquad \underline{bwcheck}var({\it string}) \underline{bwrestrict}var({\it string}) \underline{stdvars}var({\it string}) \\
    \qquad\qquad \underline{vce}var({\it string}) level({\it \#}) plot graph\_opt({\it string})}
}
\end{stsyntax}
\color{black}
\hangpara {\it depvar} is the dependent variable.

\hangpara {\it runvar1} is the running variable (a.k.a. score or forcing variable) in a cumulative cutoffs setting.

\hangpara {\it runvar2} if specified, is the second running variable (a.k.a. score or forcing variable) in a two-score setting.

\hangpara {\it treatvar} if specified, is the treatment indicator in a two-score setting.

\hangpara \texttt{\underline{c}var({\it cutoff\_var1 [cutoff\_var2]})} specifies the numeric variable \textit{cutoff\_var1} that indicates the cutoff faced by each unit in the sample in a cumulative cutoff setting, or the two running variables \textit{cutoff\_var1} and \textit{cutoff\_var2} in a two-score RD design.

\hangpara \texttt{range({\it range1 [range2]})} specifies the range of the running variable to be used for estimation around each cutoff. Specifying only one variable implies using the same range at each side of the cutoff.

\hangpara \texttt{xnorm({\it string})} specifies the normalized running variable to estimate pooled effect.

\hangpara \texttt{fuzzy({\it string})} indicates a fuzzy design. See \texttt{help rdrobust} for details.

\hangpara \texttt{\underline{deriv}var({\it string})}  a variable of length equal to the number of different cutoffs that specifies the order of the derivative for \texttt{rdrobust} to calculate cutoff-specific estimates. See \texttt{help rdrobust} for details.

\hangpara \texttt{pooled\_opt({\it string})} specifies the options to be passed to \texttt{rdrobust} to calculate pooled estimate. See \texttt{help rdrobust} for details.

\hangpara \texttt{\underline{p}var({\it string})} a variable of length equal to the number of different cutoffs that specifies the order of the polynomials for \texttt{rdrobust} to calculate cutoff-specific estimates. See \texttt{help rdrobust} for details.

\hangpara \texttt{\underline{q}var({\it string})} a variable of length equal to the number of different cutoffs that specifies the order of the polynomials for bias estimation for \texttt{rdrobust} to calculate cutoff-specific estimates. See \texttt{help rdrobust} for details.

\hangpara \texttt{\underline{h}var({\it string})} a variable of length equal to the number of different cutoffs that specifies the bandwidths for \texttt{rdrobust} to calculate cutoff-specific estimates. When \texttt{hrightvar} is specified, \texttt{hvar} indicates the bandwidth to the left of the cutoff.  When \texttt{hrightvar} is not specified, the same bandwidths are used at each side.  See \texttt{help rdrobust} for details.

\hangpara \texttt{\underline{hright}var({\it string})} a variable of length equal to the number of different cutoffs that specifies the bandwidths to the right of the cutoff for \texttt{rdrobust} to calculate cutoff-specific estimates. When \texttt{hrightvar} is not specified, the same bandwidths in \texttt{hvar} are used at each side.  See \texttt{help rdrobust} for details.

\hangpara \texttt{\underline{b}var({\it string})} a variable of length equal to the number of different cutoffs that specifies the bandwidths for the bias for \texttt{rdrobust} to calculate cutoff-specific estimates. When \texttt{brightvar} is specified, \texttt{bvar} indicates the bandwidth to the left of the cutoff.  When \texttt{brightvar} is not specified, the same bandwidths are used at each side.  See \texttt{help rdrobust} for details.

\hangpara \texttt{\underline{bright}var({\it string})} a variable of length equal to the number of different cutoffs that specifies the bandwidths to the right of the cutoff for \texttt{rdrobust} to calculate cutoff-specific estimates. When \texttt{brightvar} is not specified, the same bandwidths in \texttt{bvar} are used at each side.  See \texttt{help rdrobust} for details.

\hangpara \texttt{\underline{rho}var({\it string})}  a variable of length equal to the number of different cutoffs that specifies the value of rho for \texttt{rdrobust} to calculate cutoff-specific estimates. See \texttt{help rdrobust} for details.

\hangpara \texttt{\underline{covs}var({\it string})}  a variable of length equal to the number of different cutoffs that specifies the covariates for \texttt{rdrobust} to calculate cutoff-specific estimates. See \texttt{help rdrobust} for details.

\hangpara \texttt{\underline{covsdrop}var({\it string})} a variable of length equal to the number of different cutoffs that specifies whether collinear covariates should be dropped. See \texttt{help rdrobust} for details.

\hangpara \texttt{\underline{kernel}var({\it string})} a variable of length equal to the number of different cutoffs that specifies the kernels for \texttt{rdrobust} to calculate cutoff-specific estimates. See \texttt{help rdrobust} for details.

\hangpara \texttt{\underline{weights}var({\it string})} a variable of length equal to the number of different cutoffs that specifies the weights for \texttt{rdrobust} to calculate cutoff-specific estimates. See \texttt{help rdrobust} for details.

\hangpara \texttt{\underline{bwselect}var({\it string})} a variable of length equal to the number of different cutoffs that specifies the bandwidth selection method for \texttt{rdrobust} to calculate cutoff-specific estimates. See \texttt{help rdrobust} for details.

\hangpara \texttt{\underline{scalepar}var({\it string})} a variable of length equal to the number of different cutoffs that specifies the value of scalepar for \texttt{rdrobust} to calculate cutoff-specific estimates. See \texttt{help rdrobust} for details.

\hangpara \texttt{\underline{scaleregul}var({\it string})} a variable of length equal to the number of different cutoffs that specifies the value of scaleregul for \texttt{rdrobust} to calculate cutoff-specific estimates. See \texttt{help rdrobust} for details.

\hangpara \texttt{\underline{masspoints}var({\it string})}  a variable of length equal to the number of different cutoffs that specifies how to handle repeated values in the running variable. See \texttt{help rdrobust} for details.

\hangpara \texttt{\underline{bwcheck}var({\it string})} a variable of length equal to the number of different cutoffs that specifies the value of \texttt{bwcheck}. See \texttt{help rdrobust} for details.

\hangpara \texttt{\underline{bwrestrict}var({\it string})} variable of length equal to the number of different cutoffs that specifies whether computed bandwidths are restricted to the range of \textit{runvar}. See \texttt{help rdrobust} for details.

\hangpara \texttt{\underline{stdvars}var({\it string})} a variable of length equal to the number of different cutoffs that specifies whether \textit{depvar} and \textit{runvar} are standardized. See \texttt{help rdrobust} for details.

\hangpara \texttt{\underline{vce}var({\it string})} a variable of length equal to the number of different cutoffs that specifies the variance-covariance matrix estimation method for \texttt{rdrobust} to calculate cutoff-specific estimates. See \texttt{help rdrobust} for details.

\hangpara \texttt{level({\it \#})} specifies the confidence levels for confidence intervals. See \texttt{help rdrobust} for details.

\hangpara \texttt{plot} plots the pooled and cutoff-specific estimates and the weights given by the pooled estimate to each cutoff-specific estimate.

\hangpara \texttt{graph\_opt({\it string})} options to be passed to the graph when \texttt{plot} is specified.

\section{Illustration of Methods}\label{section:illustration}

\subsection{Non-cumulative Multiple Cutoffs}

We begin by illustrating \texttt{rdmc} using a simulated dataset, \texttt{simdata\_multic}. In this dataset, \texttt{y} is the outcome variable, \texttt{x} is the running variable, \texttt{c} is a variable indicating the cutoff that each unit in the sample faces, and \texttt{t} is a treatment indicator, corresponding in this case to units with $\texttt{x}\ge \mathtt{c}$. As shown below, there are two different cutoffs, 33 and 66, each with the sample sample size. 

{\fontsize{8}{8}\selectfont\begin{stlog}[auto]\input{code/output/rdmc_out0.log}\end{stlog}}

The basic syntax for \texttt{rdmc} is the following:

{\fontsize{8}{8}\selectfont\begin{stlog}[auto]\input{code/output/rdmc_out1.log}\end{stlog}}

The output shows the cutoff-specific estimate at each cutoff, together with the corresponding robust bias-corrected p-value, 95 percent robust confidence interval and sample size at each cutoff, and two ``global'' estimates. The first one is a weighted average of the cutoff specific estimates using the estimated weights described in Section \ref{section:methods}. These estimated weights are shown in the last column. The second one is the pooled estimate obtained by normalizing the running variable. While these two estimators converge to the same population parameter, they can differ in finite samples as seen above. In this example, the effect is statistically significant at both cutoffs.

All the results in the above display are calculated using \texttt{rdrobust}. The user can specify options for \texttt{rdrobust} to calculate the pooled estimates using the option \texttt{pooled\_opt}. For instance, the syntax below specifies a bandwidth of 20 and a local quadratic polynomial for the pooled estimand. By default, \texttt{rdmc} omits the output from \texttt{rdrobust} when estimating the effects. The output from the pooled effect estimation can be displayed using the option \texttt{verbose}, which we use below to show how the options are passed to \texttt{rdrobust}.

{\fontsize{8}{8}\selectfont\begin{stlog}[auto]\input{code/output/rdmc_out2.log}\end{stlog}}

The user can also modify the options for estimation in each specific cutoff. The following syntax shows how to manually change options for the cutoff-specific estimates by setting a bandwidth of 11 in the first cutoff and 10 in the second one.

{\fontsize{8}{8}\selectfont\begin{stlog}[auto]\input{code/output/rdmc_out3.log}\end{stlog}}

All the cutoff-specific options are passed in a similar fashion, defining a new variable of length equal to the number of cutoffs that indicates the options for each cutoff in its values. For instance, the following syntax indicates different bandwidth selection methods at each cutoff:

{\fontsize{8}{8}\selectfont\begin{stlog}[auto]\input{code/output/rdmc_out4.log}\end{stlog}}

%
%
%

The \texttt{rdmc} command saves the bias-corrected estimates and variances in the matrices \texttt{e(b)} and \texttt{e(V)}, which allows for post-estimation testing using \texttt{lincom} or \texttt{test}. For instance, to test whether the effects at the two cutoffs are the same, type:

{\fontsize{8}{8}\selectfont\begin{stlog}[auto]\input{code/output/rdmc_out6.log}\end{stlog}}

\begin{figure}
\centering
\includegraphics[scale=.7]{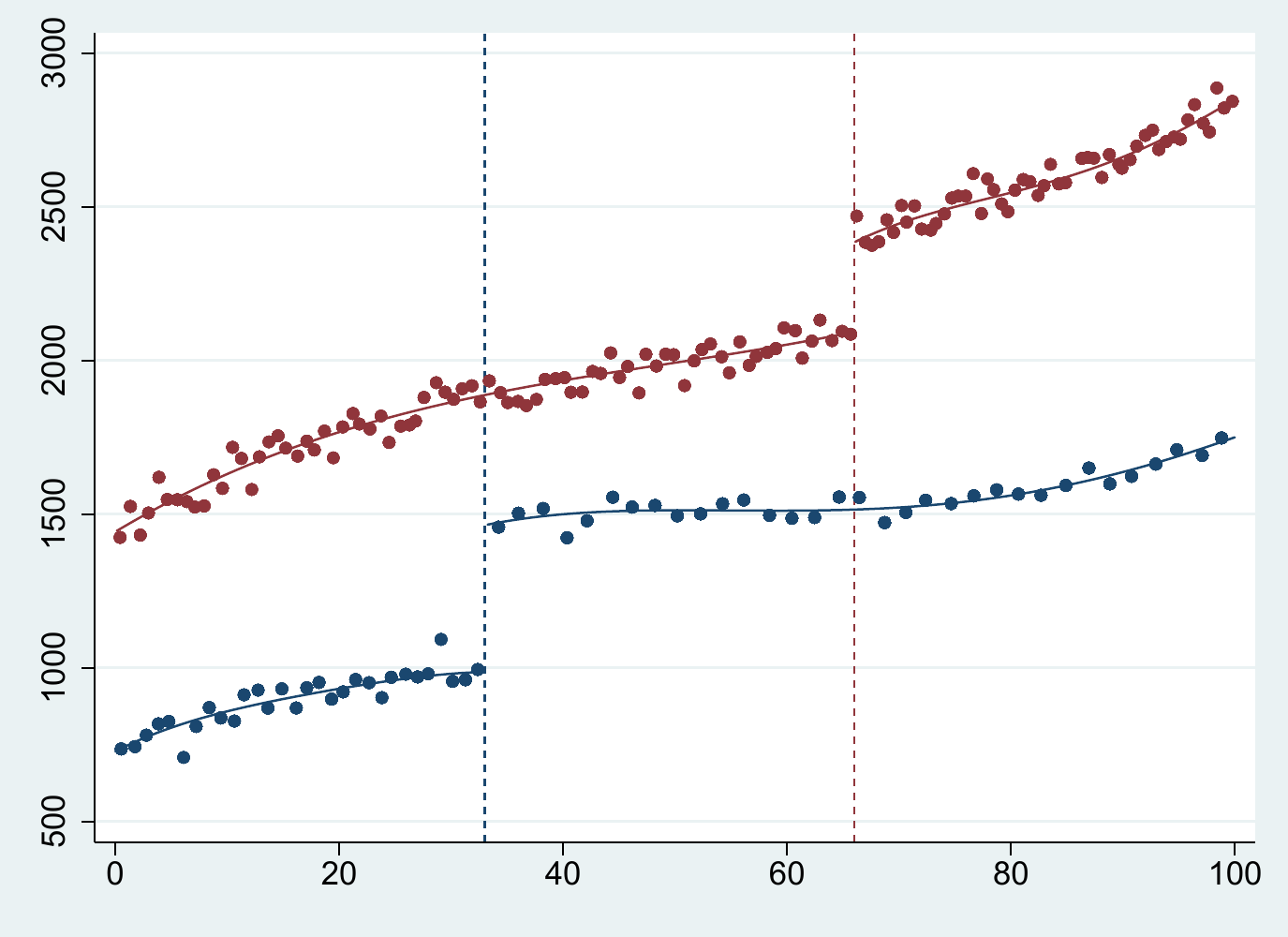}
\caption{Multiple RD plot.}\label{fig:multi}
\end{figure}

The \texttt{rdmcplot} command jointly plots the estimated regression functions at each cutoff. The output from \texttt{rdmcplot} is shown in Figure \ref{fig:multi}. The basic syntax is the following:

{\fontsize{8}{8}\selectfont\begin{stlog}[auto]\input{code/output/rdmc_out7.log}\end{stlog}}

%
%

The \texttt{rdmcplot} includes all the options available for \texttt{rdplot}. For example, the plot can be restricted to a bandwidth using the option \texttt{h()} and to use a polynomial of a specified order using the option \texttt{p()}, as shown below. This option allows the user to plot the linear fit and estimated treatment effects at each cutoff.

{\fontsize{8}{8}\selectfont\begin{stlog}[auto]\input{code/output/rdmc_out9.log}\end{stlog}}

\begin{figure}
\centering
\includegraphics[scale=.7]{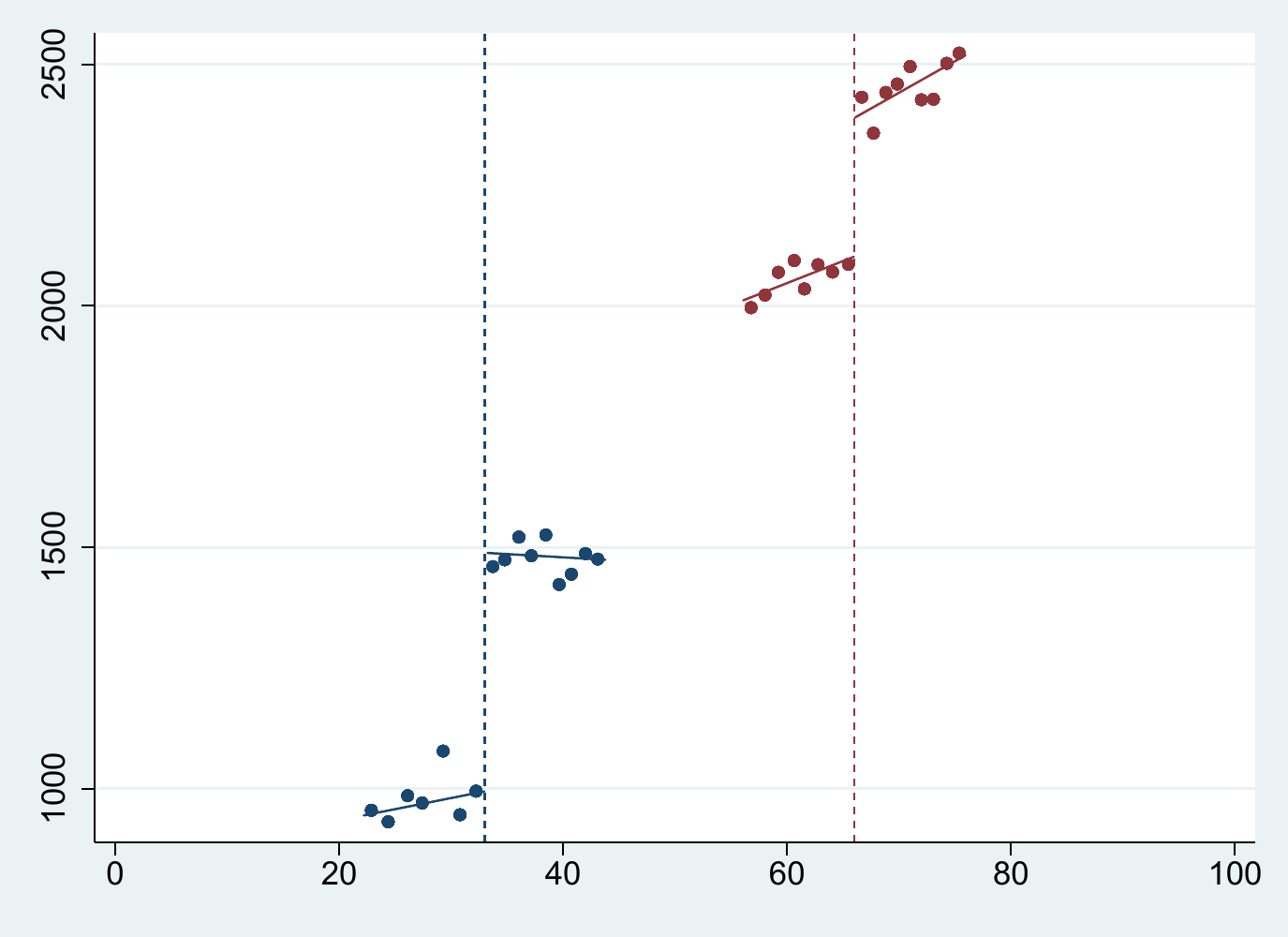}
\caption{Multiple RD plot.}\label{fig:multi3}
\end{figure}

The resulting plot is shown in Figure \ref{fig:multi3}.

The option \texttt{genvars} generates the variables required to replicate the plots by hand. This allows the user to customize the plot. The following code illustrates how to use this option to replicate Figure \ref{fig:multi3}. 

{\fontsize{8}{8}\selectfont\begin{stlog}[auto]\input{code/output/rdmc_out10.log}\end{stlog}}


\subsection{Cumulative Multiple Cutoffs}

We now illustrate the use of \texttt{rdms} for cumulative cutoffs using the simulated dataset \texttt{simdata\_cumul}. In this dataset, the running variable ranges from 0 to 100, and units with running variable below $33$ receive a certain treatment level $d_1$ whereas units with running variable above $66$ receive another treatment level $d_2$. In this setting, the cutoffs are indicated as a variable in the dataset, where each row indicates a cutoff. 

{\fontsize{8}{8}\selectfont\begin{stlog}[auto]\input{code/output/cumul_out0.log}\end{stlog}}

The syntax for cumulative cutoffs is similar to \texttt{rdmc}. The user specifies the outcome variable, the running variable and the cutoffs as follows:

{\fontsize{8}{8}\selectfont\begin{stlog}[auto]\input{code/output/cumul_out1.log}\end{stlog}}

Options like the bandwidth, polynomial order and kernel for each cutoff-specific effect can be specified by creating variables as shown below.

{\fontsize{8}{8}\selectfont\begin{stlog}[auto]\input{code/output/cumul_out2.log}\end{stlog}}

Without further information, the \texttt{rdms} command could be using any observation above the cutoff 33 to estimate the effect of the first treatment level $d_1$. This implies that some observations in the range $[66,100]$ are used. But these observations receive the second treatment level, $d_2$. This feature can result in inconsistent estimators for $\tau_1$. To avoid this problem, the user can specify the range of observations to be used around each cutoff. In this case, we can restrict the range at the first cutoff (33) to go from 0 to 65.5, to ensure that no observations above 66 are used, and the range at the second cutoff (66) to go from 33.5 to 100. This can be done as follows.

{\fontsize{8}{8}\selectfont\begin{stlog}[auto]\input{code/output/cumul_out3.log}\end{stlog}}

The pooled estimate can be obtained using \texttt{rdmc}. For this, we need to assign each unit in the sample a value for the cutoff. One possibility is to assign each unit to the closest cutoff. For this, we generate a variable named \texttt{cutoff} that equals 33 for units with score below 49.5 (the middle point between 33 and 66), and equals 66 for units above 49.5.

{\fontsize{8}{8}\selectfont\begin{stlog}[auto]\input{code/output/cumul_out4.log}\end{stlog}}

\begin{figure}
\centering
\includegraphics[scale=.7]{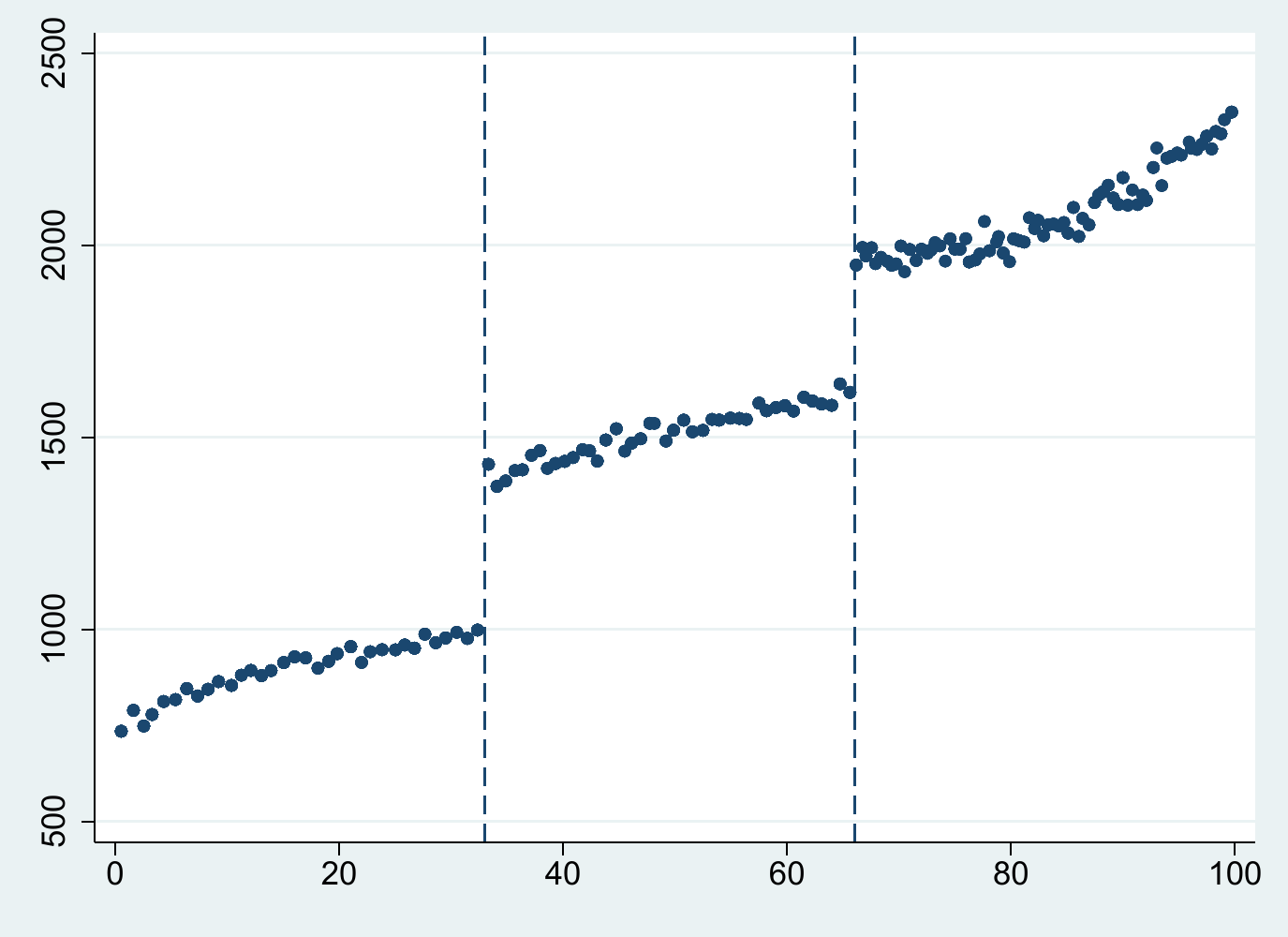}
\caption{Cumulative cutoffs.}\label{fig:cumul}
\end{figure}

Finally, we can use the variable \texttt{cutoff} to plot the regression functions using \texttt{rdmcplot}, shown in Figure \ref{fig:cumul}

{\fontsize{8}{8}\selectfont\begin{stlog}[auto]\input{code/output/cumul_out5.log}\end{stlog}}

\subsection{Multiple Scores}

We now illustrate the use of \texttt{rdms} to analyze RD designs with two running variables using the simulated dataset \texttt{simdata\_multis}. In this dataset, there are two running variables, \texttt{x1} and \texttt{x2}, ranging between 0 and 100, and units receive the treatment when $\mathtt{x1} \le 50$ and $\mathtt{x2} \le 50$. We look at three cutoffs on the boundary: (25,50), (50,50) and (50,25). 

{\fontsize{8}{8}\selectfont\begin{stlog}[auto]\input{code/output/multis_out0.log}\end{stlog}}

\begin{figure}
\centering
\includegraphics[scale=.7]{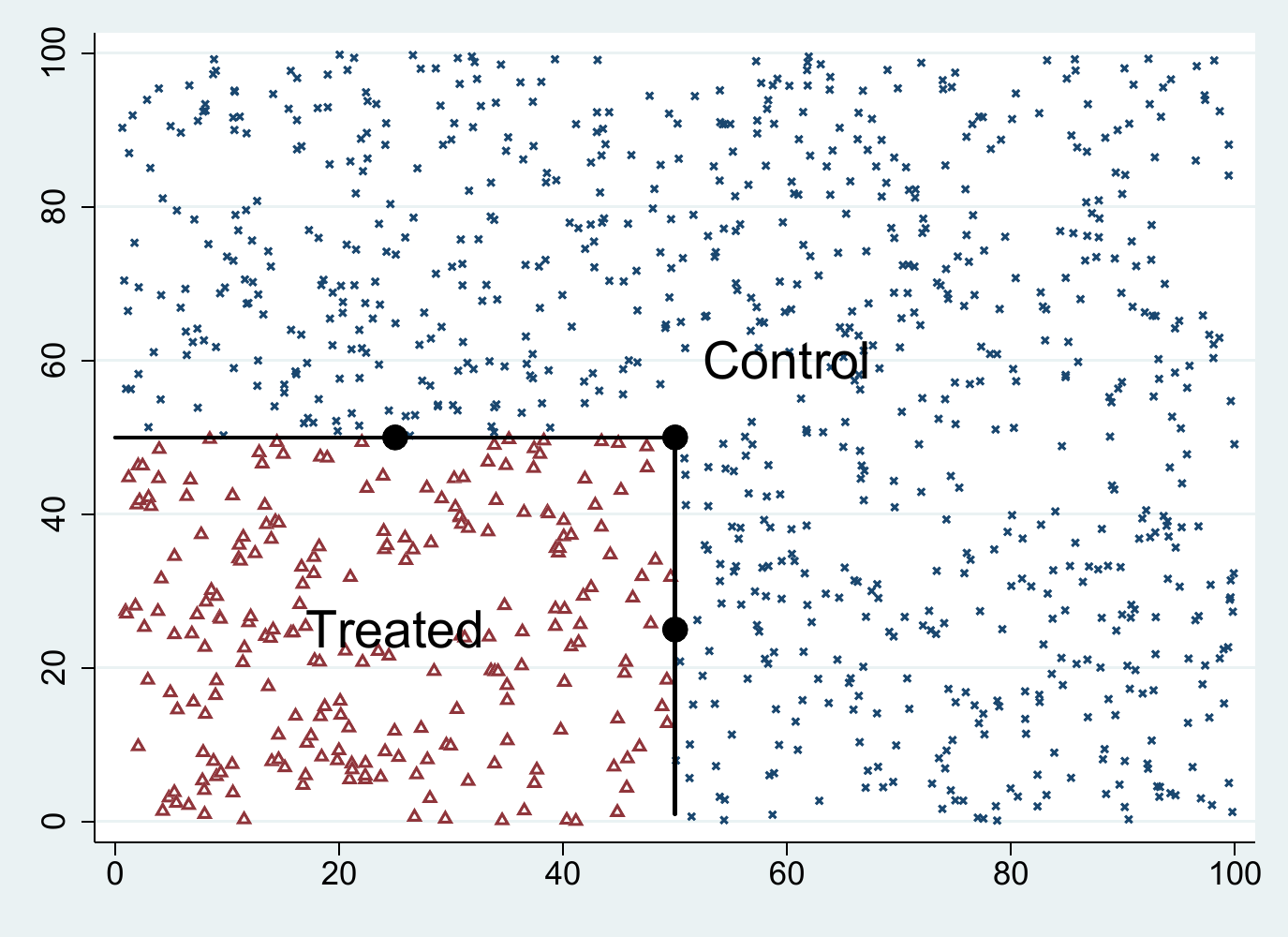}
\caption{Bivariate score.}\label{fig:multis}
\end{figure}

The following code provides a simple visualization of this setting, shown in Figure \ref{fig:multis}:

{\fontsize{8}{8}\selectfont\begin{stlog}[auto]\input{code/output/multis_out1.log}\end{stlog}}

The basic syntax is the following:

{\fontsize{8}{8}\selectfont\begin{stlog}[auto]\input{code/output/multis_out2.log}\end{stlog}}

Information to estimate each cutoff-specific estimate can be provided as illustrated before. For instance, to specify cutoff-specific bandwidths:

{\fontsize{8}{8}\selectfont\begin{stlog}[auto]\input{code/output/multis_out3.log}\end{stlog}}

Finally, the \texttt{xnorm} option allows the user to specify the normalized running variable to calculate a pooled estimate. In this case, we define the normalized running variable as the closest perpendicular distance to the boundary defined by the treatment assignment, with positive values indicating treated units and negative values indicating control units. 

{\fontsize{8}{8}\selectfont\begin{stlog}[auto]\input{code/output/multis_out4.log}\end{stlog}}

\section{Conclusion}\label{section:conclusion}

We introduced the \texttt{Stata} package \texttt{rdmulti} to anlyze RD designs with multiple cutoffs or scores. A companion \texttt{R} function with the same syntax and capabilities is also provided.

\section{Acknowledgments}

We thank Sebastian Calonico and Nicolas Idrobo for helpful comments and discussions. The authors gratefully acknowledge financial support from the National Science Foundation through grant SES-1357561.

\bibliographystyle{sj}
\bibliography{Cattaneo-Titiunik-VazquezBare_2020_Stata}

\section*{About the Authors}

\noindent
Matias D. Cattaneo is a Professor at the Department of Operations Research and Financial Engineering, Princeton University.

\noindent
Roc\'io Titiunik is a Professor of Political Science at Princeton University.

\noindent
Gonzalo Vazquez-Bare is an Assistant Professor of Economics at the University of California, Santa Barbara.

\end{document}

%% file: code/output/rdmc_out0.log.tex
. use simdata_multic, clear
{\smallskip}
. sum 
{\smallskip}
    Variable {\VBAR}        Obs        Mean    Std. Dev.       Min        Max
\HLI{13}{\PLUS}\HLI{57}
           c {\VBAR}      2,000        49.5    16.50413         33         66
           x {\VBAR}      2,000    50.79875    28.95934   .0184725   99.97507
           t {\VBAR}      2,000        .516    .4998689          0          1
           y {\VBAR}      2,000    1728.135    545.0856   540.0849   3015.232
{\smallskip}
. tab c
{\smallskip}
          c {\VBAR}      Freq.     Percent        Cum.
\HLI{12}{\PLUS}\HLI{35}
         33 {\VBAR}      1,000       50.00       50.00
         66 {\VBAR}      1,000       50.00      100.00
\HLI{12}{\PLUS}\HLI{35}
      Total {\VBAR}      2,000      100.00
{\smallskip}

%% file: code/output/rdmc_out1.log.tex
. rdmc y x, c(c)
{\smallskip}
{\smallskip}
Cutoff-specific RD estimation with robust bias-corrected inference
\HLI{12}{\TOPT}\HLI{67}
      Cutoff{\VBAR}    Coef.    P>|z|  [95\% Conf. Int.]     hl     hr      Nh   Weight
\HLI{12}{\PLUS}\HLI{67}
          33{\VBAR}  484.831    0.00    421.18  552.53    14.66  14.66    289    0.540
          66{\VBAR}  297.981    0.00    220.35  362.27    11.95  11.95    246    0.460
\HLI{12}{\PLUS}\HLI{67}
    Weighted{\VBAR}  398.915    0.00    348.74  445.14        .      .    535        .
      Pooled{\VBAR}  436.400    0.00    179.34  676.63    13.68  13.68    550        .
\HLI{12}{\BOTT}\HLI{67}

%% file: code/output/rdmc_out2.log.tex
. rdmc y x, c(c) pooled_opt(h(20) p(2)) verbose
{\smallskip}
Sharp RD estimates using local polynomial regression.
{\smallskip}
      Cutoff c = 0 {\VBAR} Left of c  Right of c            Number of obs =       2000
\HLI{19}{\PLUS}\HLI{22}            BW type       =     Manual
     Number of obs {\VBAR}       968        1032            Kernel        = Triangular
Eff. Number of obs {\VBAR}       409         416            VCE method    =         NN
    Order est. (p) {\VBAR}         2           2
    Order bias (q) {\VBAR}         3           3
       BW est. (h) {\VBAR}    20.000      20.000
       BW bias (b) {\VBAR}    20.000      20.000
         rho (h/b) {\VBAR}     1.000       1.000
{\smallskip}
Outcome: y. Running variable: __000002.
\HLI{19}{\TOPT}\HLI{60}
            Method {\VBAR}   Coef.    Std. Err.    z     P>|z|    [95\% Conf. Interval]
\HLI{19}{\PLUS}\HLI{60}
      Conventional {\VBAR}  437.04      129.8   3.3671   0.001    182.643      691.441
            Robust {\VBAR}     -          -     3.0118   0.003    185.618      877.381
\HLI{19}{\BOTT}\HLI{60}
{\smallskip}
{\smallskip}
Cutoff-specific RD estimation with robust bias-corrected inference
\HLI{12}{\TOPT}\HLI{67}
      Cutoff{\VBAR}    Coef.    P>|z|  [95\% Conf. Int.]     hl     hr      Nh   Weight
\HLI{12}{\PLUS}\HLI{67}
          33{\VBAR}  484.831    0.00    421.18  552.53    14.66  14.66    289    0.540
          66{\VBAR}  297.981    0.00    220.35  362.27    11.95  11.95    246    0.460
\HLI{12}{\PLUS}\HLI{67}
    Weighted{\VBAR}  398.915    0.00    348.74  445.14        .      .    535        .
      Pooled{\VBAR}  437.042    0.00    185.62  877.38    20.00  20.00    825        .
\HLI{12}{\BOTT}\HLI{67}

%% file: code/output/rdmc_out3.log.tex
. gen double h = 11 in 1
(1,999 missing values generated)
{\smallskip}
. replace h = 10 in 2
(1 real change made)
{\smallskip}
. rdmc y x, c(c) h(h)
{\smallskip}
{\smallskip}
Cutoff-specific RD estimation with robust bias-corrected inference
\HLI{12}{\TOPT}\HLI{67}
      Cutoff{\VBAR}    Coef.    P>|z|  [95\% Conf. Int.]     hl     hr      Nh   Weight
\HLI{12}{\PLUS}\HLI{67}
          33{\VBAR}  495.429    0.00    368.13  563.21    11.00  11.00    207    0.498
          66{\VBAR}  303.769    0.00    220.40  403.32    10.00  10.00    209    0.502
\HLI{12}{\PLUS}\HLI{67}
    Weighted{\VBAR}  399.138    0.00    321.56  455.23        .      .    416        .
      Pooled{\VBAR}  436.400    0.00    179.34  676.63    13.68  13.68    550        .
\HLI{12}{\BOTT}\HLI{67}

%% file: code/output/rdmc_out4.log.tex
. gen bwselect = "msetwo" in 1
(1,999 missing values generated)
{\smallskip}
. replace bwselect = "certwo" in 2
(1 real change made)
{\smallskip}
. rdmc y x, c(c) bwselect(bwselect)
{\smallskip}
{\smallskip}
Cutoff-specific RD estimation with robust bias-corrected inference
\HLI{12}{\TOPT}\HLI{67}
      Cutoff{\VBAR}    Coef.    P>|z|  [95\% Conf. Int.]     hl     hr      Nh   Weight
\HLI{12}{\PLUS}\HLI{67}
          33{\VBAR}  481.567    0.00    417.80  546.83    14.49  16.91    313    0.570
          66{\VBAR}  298.726    0.00    227.42  367.21    14.74   7.95    236    0.430
\HLI{12}{\PLUS}\HLI{67}
    Weighted{\VBAR}  402.969    0.00    355.30  450.28        .      .    549        .
      Pooled{\VBAR}  436.400    0.00    179.34  676.63    13.68  13.68    550        .
\HLI{12}{\BOTT}\HLI{67}

%% file: code/output/rdmc_out6.log.tex
. rdmc y x, c(c)
{\smallskip}
{\smallskip}
Cutoff-specific RD estimation with robust bias-corrected inference
\HLI{12}{\TOPT}\HLI{67}
      Cutoff{\VBAR}    Coef.    P>|z|  [95\% Conf. Int.]     hl     hr      Nh   Weight
\HLI{12}{\PLUS}\HLI{67}
          33{\VBAR}  484.831    0.00    421.18  552.53    14.66  14.66    289    0.540
          66{\VBAR}  297.981    0.00    220.35  362.27    11.95  11.95    246    0.460
\HLI{12}{\PLUS}\HLI{67}
    Weighted{\VBAR}  398.915    0.00    348.74  445.14        .      .    535        .
      Pooled{\VBAR}  436.400    0.00    179.34  676.63    13.68  13.68    550        .
\HLI{12}{\BOTT}\HLI{67}
{\smallskip}
. matlist e(b)
{\smallskip}
             {\VBAR}        c1         c2   weighted     pooled 
\HLI{13}{\PLUS}\HLI{44}
          y1 {\VBAR}  486.8578   291.3082   396.9415   427.9832 
{\smallskip}
. lincom c1-c2
{\smallskip}
 ( 1)  c1 - c2 = 0
{\smallskip}
\HLI{13}{\TOPT}\HLI{64}
             {\VBAR}      Coef.   Std. Err.      z    P>|z|     [95\% Conf. Interval]
\HLI{13}{\PLUS}\HLI{64}
         (1) {\VBAR}   195.5496    49.3309     3.96   0.000     98.86279    292.2364
\HLI{13}{\BOTT}\HLI{64}

%% file: code/output/rdmc_out7.log.tex
. rdmcplot y x, c(c)

%% file: code/output/rdmc_out9.log.tex
. gen p = 1 in 1/2
(1,998 missing values generated)
{\smallskip}
. rdmcplot y x, c(c) h(h) p(p)

%% file: code/output/rdmc_out10.log.tex
. rdmcplot y x, c(c) genvars
{\smallskip}
. twoway (scatter rdmcplot_mean_y_1 rdmcplot_mean_x_1, mcolor(navy)) ///
>         (line rdmcplot_hat_y_1 rdmcplot_mean_x_1 if t==1, sort lcolor(navy)) ///
>         (line rdmcplot_hat_y_1 rdmcplot_mean_x_1 if t==0, sort lcolor(navy)) ///
>         (scatter rdmcplot_mean_y_2 rdmcplot_mean_x_2, mcolor(maroon)) ///
>         (line rdmcplot_hat_y_2 rdmcplot_mean_x_2 if t==1, sort lcolor(maroon)) ///
>         (line rdmcplot_hat_y_2 rdmcplot_mean_x_2 if t==0, sort lcolor(maroon)), ///
>         xline(33, lcolor(navy) lpattern(dash)) ///
>         xline(66, lcolor(maroon) lpattern(dash)) ///
>         legend(off) 

%% file: code/output/cumul_out0.log.tex
. use simdata_cumul, clear
{\smallskip}
. sum 
{\smallskip}
    Variable {\VBAR}        Obs        Mean    Std. Dev.       Min        Max
\HLI{13}{\PLUS}\HLI{57}
           x {\VBAR}      1,000    50.46639    28.69369   .0413166    99.8783
           y {\VBAR}      1,000    1508.638    488.2752   658.4198   2480.568
           c {\VBAR}          2        49.5    23.33452         33         66
{\smallskip}
. tab c
{\smallskip}
          c {\VBAR}      Freq.     Percent        Cum.
\HLI{12}{\PLUS}\HLI{35}
         33 {\VBAR}          1       50.00       50.00
         66 {\VBAR}          1       50.00      100.00
\HLI{12}{\PLUS}\HLI{35}
      Total {\VBAR}          2      100.00
{\smallskip}

%% file: code/output/cumul_out1.log.tex
. rdms y x, c(c)
{\smallskip}
{\smallskip}
Cutoff-specific RD estimation with robust bias-corrected inference
\HLI{15}{\TOPT}\HLI{64}
         Cutoff{\VBAR}      Coef.     P>|z|     [95\% Conf. Int.]     hl     hr      Nh
\HLI{15}{\PLUS}\HLI{64}
             33{\VBAR}    395.492     0.000     363.76   423.86    15.11  15.11    286
             66{\VBAR}    342.872     0.000     315.95   373.96    12.22  12.22    265
\HLI{15}{\BOTT}\HLI{64}

%% file: code/output/cumul_out2.log.tex
. gen double h = 11 in 1
(999 missing values generated)
{\smallskip}
. replace h = 8 in 2
(1 real change made)
{\smallskip}
. gen kernel = "uniform" in 1
(999 missing values generated)
{\smallskip}
. replace kernel = "triangular" in 2
variable {\bftt{kernel}} was {\bftt{str7}} now {\bftt{str10}}
(1 real change made)
{\smallskip}
. rdms y x, c(c) h(h) kernel(kernel)
{\smallskip}
{\smallskip}
Cutoff-specific RD estimation with robust bias-corrected inference
\HLI{15}{\TOPT}\HLI{64}
         Cutoff{\VBAR}      Coef.     P>|z|     [95\% Conf. Int.]     hl     hr      Nh
\HLI{15}{\PLUS}\HLI{64}
             33{\VBAR}    394.470     0.000     351.65   438.72    11.00  11.00    215
             66{\VBAR}    342.505     0.000     301.56   375.95     8.00   8.00    166
\HLI{15}{\BOTT}\HLI{64}

%% file: code/output/cumul_out3.log.tex
. gen double range_l = 0 in 1
(999 missing values generated)
{\smallskip}
. gen double range_r = 65.5 in 1
(999 missing values generated)
{\smallskip}
. replace range_l = 33.5 in 2
(1 real change made)
{\smallskip}
. replace range_r = 100 in 2
(1 real change made)
{\smallskip}
. rdms y x, c(c) range(range_l range_r)
{\smallskip}
{\smallskip}
Cutoff-specific RD estimation with robust bias-corrected inference
\HLI{15}{\TOPT}\HLI{64}
         Cutoff{\VBAR}      Coef.     P>|z|     [95\% Conf. Int.]     hl     hr      Nh
\HLI{15}{\PLUS}\HLI{64}
             33{\VBAR}    394.698     0.000     356.12   430.45    10.96  10.96    214
             66{\VBAR}    342.180     0.000     312.20   372.04    11.18  11.18    246
\HLI{15}{\BOTT}\HLI{64}

%% file: code/output/cumul_out4.log.tex
. gen double cutoff = c[1]*(x<=49.5) + c[2]*(x>49.5)
{\smallskip}
. rdmc y x, c(cutoff)
{\smallskip}
{\smallskip}
Cutoff-specific RD estimation with robust bias-corrected inference
\HLI{12}{\TOPT}\HLI{67}
      Cutoff{\VBAR}    Coef.    P>|z|  [95\% Conf. Int.]     hl     hr      Nh   Weight
\HLI{12}{\PLUS}\HLI{67}
          33{\VBAR}  389.528    0.00    332.94  443.69     6.26   6.26    119    0.531
          66{\VBAR}  341.015    0.00    300.39  377.33     5.04   5.04    105    0.469
\HLI{12}{\PLUS}\HLI{67}
    Weighted{\VBAR}  366.788    0.00    330.63  399.64        .      .    224        .
      Pooled{\VBAR}  363.968    0.00    180.11  551.78     8.14   8.14    333        .
\HLI{12}{\BOTT}\HLI{67}

%% file: code/output/cumul_out5.log.tex
. gen binsopt = "mcolor(navy)" in 1/2
(998 missing values generated)
{\smallskip}
. gen xlineopt = "lcolor(navy) lpattern(dash)" in 1/2
(998 missing values generated)
{\smallskip}
. rdmcplot y x, c(cutoff) binsoptvar(binsopt) xlineopt(xlineopt) nopoly

%% file: code/output/multis_out0.log.tex
. use simdata_multis, clear
{\smallskip}
. sum 
{\smallskip}
    Variable {\VBAR}        Obs        Mean    Std. Dev.       Min        Max
\HLI{13}{\PLUS}\HLI{57}
          x1 {\VBAR}      1,000    50.22881    28.87868   .6323666   99.94879
          x2 {\VBAR}      1,000    50.63572     29.1905   .0775479   99.78458
           t {\VBAR}      1,000        .223    .4164666          0          1
           y {\VBAR}      1,000    728.5048    205.5627   329.4558   1372.777
          c1 {\VBAR}          3    41.66667    14.43376         25         50
\HLI{13}{\PLUS}\HLI{57}
          c2 {\VBAR}          3    41.66667    14.43376         25         50
{\smallskip}
. list c1 c2 in 1/3
{\smallskip}
     {\TLC}\HLI{9}{\TRC}
     {\VBAR} c1   c2 {\VBAR}
     {\LFTT}\HLI{9}{\RGTT}
  1. {\VBAR} 25   50 {\VBAR}
  2. {\VBAR} 50   50 {\VBAR}
  3. {\VBAR} 50   25 {\VBAR}
     {\BLC}\HLI{9}{\BRC}

%% file: code/output/multis_out1.log.tex
. gen xaux = 50 in 1/50
(950 missing values generated)
{\smallskip}
. gen yaux = _n in 1/50
(950 missing values generated)
{\smallskip}
. twoway (scatter x2 x1 if t==0, msize(small) mfcolor(white) msymbol(X)) ///
>            (scatter x2 x1 if t==1, msize(small) mfcolor(white) msymbol(T)) ///
>            (function y = 50, range(0 50) lcolor(black) lwidth(medthick)) ///
>            (line yaux xaux, lcolor(black) lwidth(medthick)) ///
>            (scatteri 50 25, msize(large) mcolor(black)) ///
>            (scatteri 50 50, msize(large) mcolor(black)) ///
>            (scatteri 25 50, msize(large) mcolor(black)), ///
>            text(25 25 "Treated", size(vlarge)) ///
>            text(60 60 "Control", size(vlarge)) ///
>            legend(off)

%% file: code/output/multis_out2.log.tex
. rdms y x1 x2 t, c(c1 c2)
{\smallskip}
{\smallskip}
Cutoff-specific RD estimation with robust bias-corrected inference
\HLI{15}{\TOPT}\HLI{64}
         Cutoff{\VBAR}      Coef.     P>|z|     [95\% Conf. Int.]     hl     hr      Nh
\HLI{15}{\PLUS}\HLI{64}
        (25,50){\VBAR}    243.842     0.111     -50.93   491.18    11.12  11.12     42
        (50,50){\VBAR}    578.691     0.000     410.83   764.88    13.83  13.83     47
        (50,25){\VBAR}    722.444     0.000     451.49  1060.15    10.83  10.83     38
\HLI{15}{\BOTT}\HLI{64}

%% file: code/output/multis_out3.log.tex
. gen double h = 15 in 1
(999 missing values generated)
{\smallskip}
. replace h = 13 in 2
(1 real change made)
{\smallskip}
. replace h = 17 in 3
(1 real change made)
{\smallskip}
. rdms y x1 x2 t, c(c1 c2) h(h)
{\smallskip}
{\smallskip}
Cutoff-specific RD estimation with robust bias-corrected inference
\HLI{15}{\TOPT}\HLI{64}
         Cutoff{\VBAR}      Coef.     P>|z|     [95\% Conf. Int.]     hl     hr      Nh
\HLI{15}{\PLUS}\HLI{64}
        (25,50){\VBAR}    336.121     0.233    -119.35   491.36    15.00  15.00     87
        (50,50){\VBAR}    583.047     0.000     501.94  1101.24    13.00  13.00     42
        (50,25){\VBAR}    620.692     0.000     464.92  1159.99    17.00  17.00     86
\HLI{15}{\BOTT}\HLI{64}

%% file: code/output/multis_out4.log.tex
. gen double aux1 = abs(50 - x1)
{\smallskip}
. gen double aux2 = abs(50 - x2)
{\smallskip}
. egen xnorm = rowmin(aux1 aux2)
{\smallskip}
. replace xnorm = xnorm*(2*t-1)
(777 real changes made)
{\smallskip}
. rdms y x1 x2 t, c(c1 c2) xnorm(xnorm)
{\smallskip}
{\smallskip}
Cutoff-specific RD estimation with robust bias-corrected inference
\HLI{15}{\TOPT}\HLI{64}
         Cutoff{\VBAR}      Coef.     P>|z|     [95\% Conf. Int.]     hl     hr      Nh
\HLI{15}{\PLUS}\HLI{64}
        (25,50){\VBAR}    243.842     0.111     -50.93   491.18    11.12  11.12     42
        (50,50){\VBAR}    578.691     0.000     410.83   764.88    13.83  13.83     47
        (50,25){\VBAR}    722.444     0.000     451.49  1060.15    10.83  10.83     38
\HLI{15}{\PLUS}\HLI{64}
         Pooled{\VBAR}    447.017     0.000     389.33   496.85    12.73  12.73    433
\HLI{15}{\BOTT}\HLI{64}